\documentclass{article}

\usepackage{arxiv}

\usepackage[utf8]{inputenc} % allow utf-8 input
\usepackage[T1]{fontenc}    % use 8-bit T1 fonts
\usepackage{hyperref}       % hyperlinks
\usepackage{url}            % simple URL typesetting
\usepackage{booktabs}       % professional-quality tables
\usepackage{amsfonts}       % blackboard math symbols
\usepackage{nicefrac}       % compact symbols for 1/2, etc.
\usepackage{microtype}      % microtypography
\usepackage{graphicx}       % microtypography

\title{Evaluating one-shot tournament predictions}

\usepackage{amsthm}
\usepackage{amsmath}
\usepackage{amssymb}

\usepackage{natbib}% Citation support using natbib.sty
\bibpunct[, ]{(}{)}{;}{a}{}{,}% Citation support using natbib.sty
\theoremstyle{plain}% Theorem-like structures provided by amsthm.sty

\theoremstyle{definition}

\theoremstyle{remark}

\usepackage{hyperref}
\usepackage[utf8]{inputenc}
\usepackage{bbm}
\usepackage{multirow}
\usepackage[table]{xcolor}
\def\tightlist{}

\author{Claus Thorn Ekstrøm\thanks{To whom correspondance should be addressed}\\
Section of Biostatistics\\ Department of Public Health\\ University of Copenhagen, Denmark \\
  \texttt{ekstrom@sund.ku.dk} \\
   \And
  Hans Van Eetvelde \\
  Department of Applied Mathematics\\ Computer Science and Statistics\\ Universiteit Gent, Belgium \\
  \texttt{Hans.VanEetvelde@UGent.be} \\
   \And
  Christophe Ley \\
  Department of Applied Mathematics\\ Computer Science and Statistics\\ Universiteit Gent, Belgium \\
  \texttt{Christophe.Ley@UGent.be} \\
   \And
  Ulf Brefeld \\
  Machine Learning Group\\ Leuphana University of L\"uneburg, Germany \\
  \texttt{brefeld@leuphana.de} 
}

\begin{document}

\maketitle

% \maketitle

\begin{abstract}
We introduce the Tournament Rank Probability Score (TRPS) as a measure to evaluate and compare pre-tournament predictions, where predictions of the full tournament results are required to be available before the tournament begins. The TRPS handles partial ranking of teams, gives credit to predictions that are only slightly wrong, and can be modified with weights to stress the importance of particular features of the tournament prediction. Thus, the Tournament Rank Prediction Score is more flexible than the commonly preferred log loss score for such tasks. In addition, we show how predictions from historic tournaments can be optimally combined into ensemble predictions in order to maximize the TRPS for a new tournament.
\end{abstract}


\hypertarget{introduction}{%
\section{Introduction}\label{introduction}}

Predicting the winner of a sports tournament has become an ever increasing challenge for researchers, sports fans and the growing business of bookmakers. Before the start of major tournaments such as the FIFA World Cup 2018, Rugby World Cup, or Wimbledon 2019, the world press are debating, comparing and summarising the various tournament predictions and the quality of these predictions is evaluated and scrutinised after the tournaments end.

Several recent research papers have been published that predict the outcome of full tournaments in football, basketball, and tennis involving techniques from statistics \citep[\citet{Neud:Ross:2018}, \citet{Gu2019}]{groll2018}, machine learning \citep{Huang:2011} and operational research \citep{dyte2000}, respectively. The overall quality of these full tournament predictions is usually evaluated on an overly simplistic basis: how well was the actual tournament winner ranked, or how well did the highest predicted teams perform in the end? Whilst natural at first glance, such comparisons are very limited: if, for example, a prediction stated that Germany was the most likely country to win the 2018 World Cup with a probability of 13.5\% then this will be seen as a bad prediction given the fact that Germany left the tournament in the first round. However, the same prediction also reflects that Germany was predicted \emph{not} to win the tournament with probability 86.5\% which puts a lot of confidence on the actual outcome of the tournament. Focussing solely on the prediction of the winner of the tournament is too restrictive a viewpoint for evaluation and comparison of full tournament predictions.

There exist good-practice ways to compare tournament predictions on the basis of individual matches, allowing for prediction updates after every played game/round of the tournament.
For example, many websites launched online competitions for the FIFA World Cup 2018\footnote{E.g., \href{https://fifaexperts.com}{fifaexperts.com}} where participants could enter the chances of winning, drawing and losing for every single match. Based on a scoring rule, participants earned points for their educated guesses after every match.
Similarly, several research papers compare different prediction methods on a match basis by using the log-loss or the rank probability score (RPS) as loss functions \citep[\citet{baboota}, \citet{hubhub}]{ConsFent2012}. However, these approaches are not entirely satisfactory, as they allow predictions to be updated as the tournament progresses, and completely wrong initial predictions such as that of Germany as the winner in the WC 2018 will be diluted in this process. To our knowledge, there does not exist a scientifically sound way to compare predictions made before the start of a big tournament to its final outcome.

In this paper, we intend to fill this gap. More precisely, we introduce the Tournament Rank Probability Score (TRPS) as a tournament prediction performance measure as a way to evaluate the quality of each prediction by comparison to the observed final ranking, and the best prediction is the one that most closely resembled the outcome of the full tournament.
An interesting consequence of the novel TRPS is the second main goal of this paper, namely an ensemble prediction strategy where, based on previous tournaments, we optimally combine tournament predictions based on different analytical strategies by assigning weights to each prediction in such a way that the TRPS becomes maximal.

The paper is organized as follows. In Section \ref{evaluating-tournament-predictions} we set up the notation, present the TRPS and discuss why it is a natural choice for evaluating tournament predictions. Section \ref{simulations} shows extensive simulations that illustrate the behaviour of the TRPS and how it is influenced by different tournament systems.
In Section \ref{improving-predictions-using-ensemble-predictions}, we outline and discuss how the TRPS can be used to construct ensemble predictions that improve the overall predictions. In Section \ref{worldcup} we apply the proposed score approach to evaluate and compare different predictions from the 2018 FIFA World Cup and combine earlier predictions to create a simple ensemble prediction model. The proposed tournament rank probability score and the data used for the example have been implemented as an R package \texttt{socceR} which can be found at CRAN.

\hypertarget{evaluating-tournament-predictions}{%
\section{Evaluating tournament predictions}\label{evaluating-tournament-predictions}}

\hypertarget{trps}{%
\subsection{The Tournament Rank Probability Score}\label{trps}}

Let \(X\in \mathbb{R}^{R\times T}\) represent a tournament prediction where \(T\) teams (or players) are able to attain \(R\leq T\) possible ranks. The elements of the matrix \(X\) contain the prediction probabilities that a team (the columns in the matrix) will obtain a given rank. Each element in \(X\) is a conditional probability and is therefore restricted to \([0, 1]\), and by construction each of the columns sum to 1, i.e., \(\sum_{r=1}^R X_{rt} = 1\) for all \(t\in\{1, \ldots, T\}\). Without loss of generality we assume that the ranks are ordered such that \(r=1\) is the best rank (first place winner) while \(r=R\) is the worst rank.

If individual rankings of all teams are possible then the number of ranks equals the number of teams, \(R=T\), but many tournaments only provide proper rankings of the teams at the top of the score board (e.g., first place, second place, etc.) while only partial ranking is available further down the score board (e.g., quarter finalist, 8th finalist, etc.). The tournament prediction, \(X\), should fulfill that the row sum of each rank corresponds exactly to the number of teams that will end up being assigned the given rank. For example, in the recent 2018 FIFA World Cup, there were 32 teams, and the seven possible ranks were 1st place, 2nd place, 3rd place, 4th place, 5th-8th place, 9th-16th place, and 17th-32nd place (the group stage) each containing 1, 1, 1, 1, 4, 8, and 16 teams, respectively.

When evaluating the quality of a tournament prediction we wish to take the distance between the ranks into account; we want to reward a prediction if it predicts a high probability of, say, a team obtaining rank 1 and the team in reality ranks second while penalising a prediction if it provides a high prediction probability of rank 1 and the team ranks 6th.

The Rank Probability Score (RPS) is a proper scoring rule that preserves the ordering of the ranks and places smaller penalty on predictions that are closer to the observed data than predictions that are further away from the observed data \citep{Epst1969, murphy, GneiRaft2007, ConsFent2012}. The RPS has mostly been used to evaluate the outcome of single matches with multiple ordered outcomes such as win, draw, loss, and it is defined for a single match as

\begin{equation}
\text{RPS}=\frac{1}{R-1}\sum_{r=1}^{R-1}\left(\sum_{j=1}^r (o_j - x_j)\right)^2, \label{eq:RPS}
\end{equation}

\noindent where \(R\) is the number of possible outcomes (the possible ranks of the match result), \(o_j\) is the empirical probability of outcome \(j\) (thus \(o_j\) attains the value 1 if the match resulted in outcome \(j\) and 0 otherwise), while \(x_j\) is the forecasted probability of outcome \(j\). The RPS is similar to the Brier score, but measures accuracy of a prediction when there are more than two ordered categories by using the cumulative predictions in order to be sensitive to the distance. The RPS is both non-local and sensitive to distance, so it uses all prediction probabilities and not just the one corresponding to the observed outcome and it takes ``adjacent'' predictions into account and as such it is reminiscent of the Kolmogorov-Smirnov test for comparing two cumulative distributions \citep{kolmo}. We wish to extend and adapt the rank probability score to accommodate full tournament predictions.

Let \(\mathcal{X}_{rt} = \sum_{i=1}^r X_{it}\) be the cumulative prediction that team \(t\) will obtain \emph{at least} rank \(r\) and let \(\mathcal{O}_{rt}\) be the corresponding empirical cumulative distribution function. \(\mathcal{O}_{rt}\) is readily available after the tournament ends and only attains two different values, 0 and 1: a column \(t\) in \(\mathcal{O}_{rt}\) is 0 until the rank which team \(t\) obtained in the tournament, after which point it is 1. Figure 1 shows an example where the cumulative prediction probabilities and the empirical cumulative prediction are compared.

\begin{figure}
\includegraphics[width=0.8\linewidth]{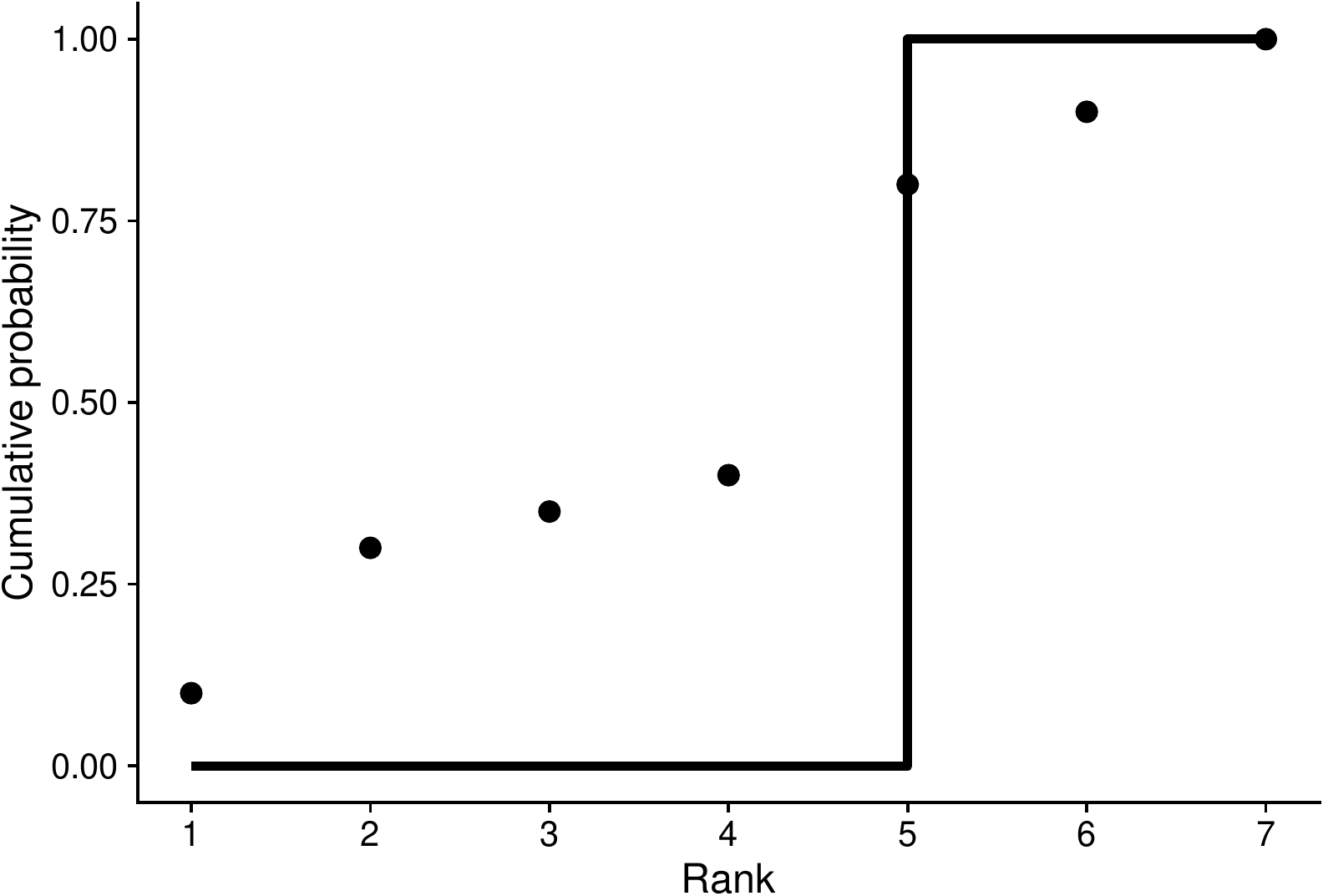} \caption{Comparison of cumulative prediction probabilities of obtaining at least rank $k$ (shown by the circles) and the actual observed cumulative prediction function (shown by the step function). Here, the team in question obtained rank 5 which is why the empirical distribution jumps for that rank.}\label{fig:unnamed-chunk-1}
\end{figure}

We define the tournament rank probability score (TRPS) for a full tournament prediction, \(X\), with actual outcome \(O\) as

\begin{equation}
\text{TRPS}(O, X) = \frac{1}{T}\sum_{t=1}^{T}\frac{1}{R-1}\sum_{r=1}^{R-1} (\mathcal{O}_{rt} - \mathcal{X}_{rt})^2. \label{eq:TRPS}
\end{equation}

A perfect prediction will result in a TRPS of 0 while the TRPS increases when the prediction becomes worse. In the definition \eqref{eq:TRPS}, we follow the definition of \eqref{eq:RPS} and only sum to \(R-1\) since \(\mathcal{O}_{Rt} - \mathcal{X}_{Rt} = 0\) for all teams because the probability that each team will obtain at least the worst ranking, \(R\), will be 1.

Note that the TRPS \eqref{eq:TRPS} - just like the RPS - utilizes the ordering of the ranks through the cumulative distribution function. This ensures that the measure is sensitive to rank distance, since the individual cumulative prediction probability comparisons at each rank are oblivious to other parts of the distribution. Thus, we also implicitly assume equal distances between the ranks when computing the score. This can be improved by adding weights to the individual ranks which also allows putting greater weights on specific predictions, for example if it is especially important to determine the winner (see Section \ref{weighted-rps}).

Figure 2 shows the behaviour of the TRPS for completely flat predictions, i.e., random guessing where each team is assigned equal probability of each ranks. When each team can be assigned a unique rank then the TRPS converges to a level around 0.17.
This level decreases steadily as a function of the number of teams when only partial ranks are possible. When a large number of teams are grouped into a single (partial) rank then it becomes easier to obtain a lower TRPS since most teams will be correctly assigned to the rank comprising many teams. In practice, any real tournament prediction should reach a value \emph{lower} than the flat TRPS in order to provide some additional information about the outcome of the tournament.

\begin{figure}
\includegraphics[width=0.8\linewidth]{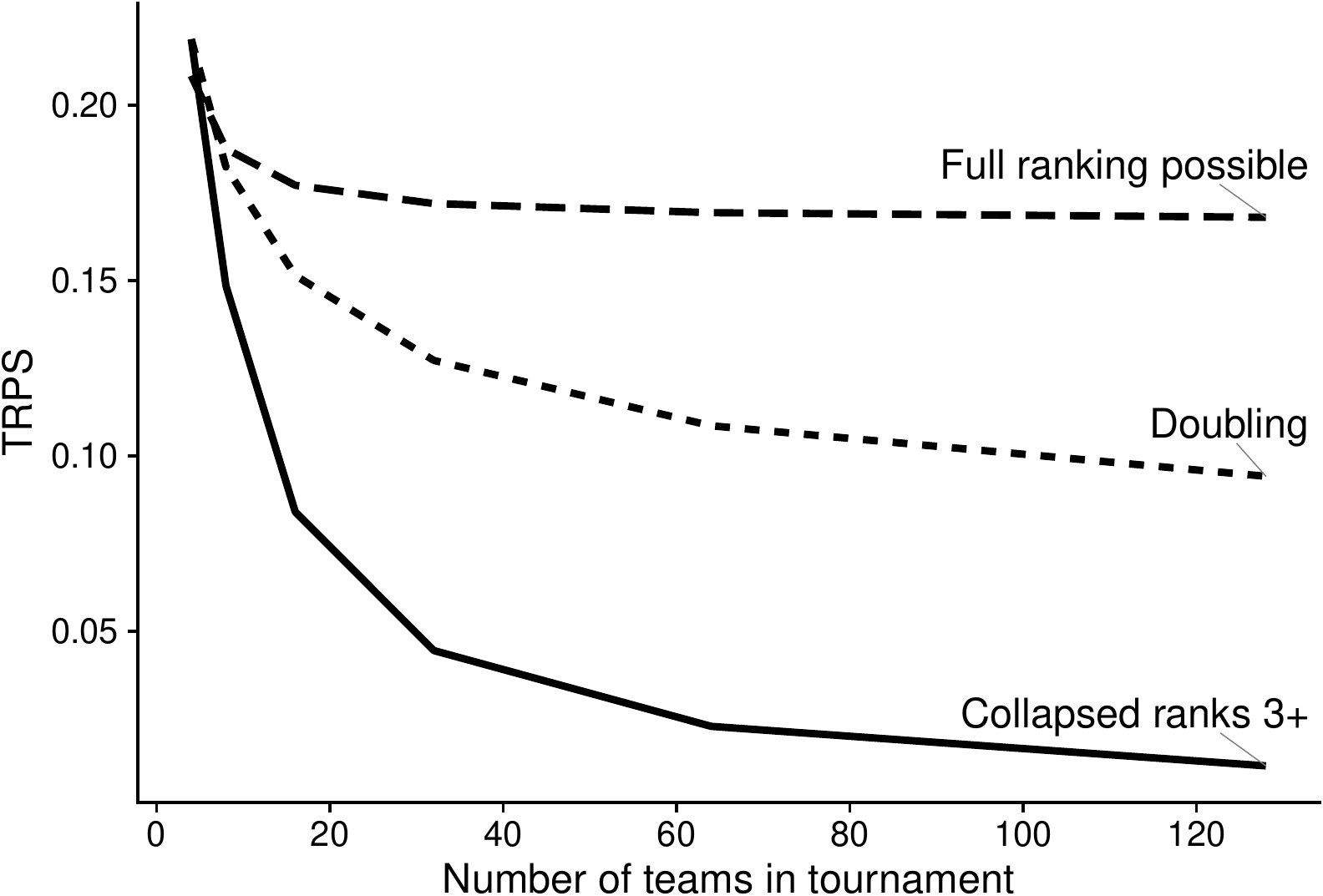} \caption{TRPS for flat predictions, i.e., where each team has the same probability of obtaining each rank. The dashed line shows the TRPS when a full ranking is available for all teams, the solid line shows the TRPS when only the first two teams can be fully ranked, and the remaining teams are grouped together into a single rank corresponding to "third or higher place", and the dotted line shows the TRPS for a knockout tournament, where each rank size is doubled (1st place, 2nd place, 3rd-4th, 5th-8th etc).}\label{fig:unnamed-chunk-2}
\end{figure}

\hypertarget{examples}{%
\subsection{Examples}\label{examples}}

\textbf{Example 1}. Consider a tournament with 4 teams that are ranked into 3 categories: 1st place, 2nd place, and 3rd-4th place. Two predictions are available prior to the start of the tournament as shown by the matrices \(X^1\) and \(X^2\) below. Note that for each prediction matrix the columns sum to 1 and the rows sum to the number of teams that will obtain each rank.

\[X^1 = \left[\begin{array}{cccc}1 & 0 & 0 & 0 \\ 0 & 1 & 0 & 0 \\ 0 & 0 & 1 & 1 \end{array}\right] \; \; X^2 = \left[\begin{array}{cccc}0.7 & 0.1 & 0.1 & 0.1 \\ 0.1 & 0.5 & 0.2 & 0.2 \\ 0.2 & 0.4 & 0.7 & 0.7 \end{array}\right]\]

The first prediction, \(X^1\), is confident that team 1 (first column) will obtain rank 1, team 2 will get 2nd place and the remaining two teams will fill out the final rank. Prediction \(X^2\) leans towards the same teams as the first predictor but is not quite as confident about the specific rankings: it assumes that team 1 will rank 1st with probability 70\% and that team 2 will get rank 3 with probability 40\%, and so on.

If the outcome of the tournament was that team 1 came first and team 2 came second then the TRPS for predictions \(X^1\) and \(X^2\) would be 0 and 0.063, respectively. \(X^1\) made a perfect prediction and obtained a score of 0 while the uncertainty in the \(X^2\) prediction increased the TRPS slightly. A completely flat prediction would in this case have resulted in a TRPS of 0.219.

Had the outcome of the tournament instead been that team 2 came first and team 1 came second then the TRPS of \(X^1\) and \(X^2\) would be 0.25 and 0.213, respectively, and prediction \(X^2\) would obtain the best TRPS. Despite not getting the prediction entirely correct, prediction \(X^2\) still fared better because it was not so confident about a wrong prediction as \(X^1\) was. In fact, prediction \(X^1\) is so poor (because it is both confident and wrong) that it is worse than just random guessing.

\textbf{Example 2}. Here we consider two predictions representing a fairly confident prediction that teams 1 and 2 will rank 1st and 2nd (\(X^3\)), and a completely flat prediction where each team has equal probability of obtaining each rank (\(X^4\)).\}

\[X^3 = \left[\begin{array}{cccc}0.75 & 0.25 & 0 & 0 \\ 0.25 & 0.75 & 0 & 0 \\ 0 & 0 & 0.75 & 0.25\\ 0 & 0 & 0.25 & 0.75 \end{array}\right] \; \; X^4 = \left[\begin{array}{cccc}0.25 & 0.25 & 0.25 & 0.25 \\ 0.25 & 0.25 & 0.25 & 0.25 \\ 0.25 & 0.25 & 0.25 & 0.25 \\ 0.25 & 0.25 & 0.25 & 0.25\end{array}\right]\]

If the actual outcome of the tournament is that teams 1-4 rank 1-4 respectively, then the TRPS becomes 0.0208 and 0.2083 for \(X^3\) and \(X^4\),
respectively. Had the ranking of the teams been 2, 1, 4, and 3 then \(X^3\) and \(X^4\) would result in TRPS of 0.188 and 0.2083, respectively. Thus
prediction \(X^3\) still performs better than the flat prediction because of the non-locality of the TRPS. Had we instead used the log loss - which is local and
only uses the prediction probability exactly matching the outcome - then we would find that \(X^3\) obtained the same log loss as the completely flat prediction \(X^4\) despite being able to identify that teams 1 and 2 should rank at the top, and teams 3 and 4 should rank at the bottom.

\hypertarget{weighted-rps}{%
\subsection{Weighted TRPS}\label{weighted-rps}}

In some instances it may be of particular interest to put emphasis on certain parts of the prediction, e.g., if it is only of interest to find the winner, or if it only is of interest to evaluate the prediction of which teams will make it past the group stages in a tournament. The TRPS can be adjusted with weights to stress the importance of particular rank predictions.

Let \(w = (w_1, \ldots, w_{R-1})\) be a vector of non-negative weights --- one for each rank except the last --- that sum to \(R-1\). We shall define the weighted TRPS as

\begin{equation}
\text{wTRPS}(O, X) = \frac{1}{T}\sum_{t=1}^{T}\frac{1}{R-1}\sum_{r=1}^{R-1} w_r (\mathcal{O}_{rt} - \mathcal{X}_{rt})^2, \label{eq:wRPS}
\end{equation}

\noindent where the assumption that \(\sum_{r=1}^{R-1} w_r = R-1\) is made to ensure that the scale of the weighted TRPS is comparable to the standard TRPS \eqref{eq:TRPS}.
Note that it is not necessary to specify a weight for the last rank, \(R\), since all teams will at least reach the last rank, and the TRPS is a special case of the wTRPS where the weights are all set to one.

Weight \(w_i\) should be interpreted as the (relative) importance of predicting the teams that reach \emph{at least} rank \(i\). For example, setting \(w=(R-1, 0, 0, \ldots, 0)\) ensures that we only consider the prediction of the winner to be important, while weights \(w=(0, \ldots, 0, R-1)\) place all the emphasis on predicting the teams that make it past the initial group stage without considering their final ranks.

Often, it is more important and relevant in a tournament to be able to predict the winner and top ranks than being able to predict the lower ranks. The emphasis on predicting the top ranking teams may be stressed by modifying the weights of the top ranks. For tournaments with partial rankings we can give equal weight to each rank rather than each team
by using the inverse of the number of teams in each rank category as a \emph{relative} weight. In the definition of the wTRPS we sum over each team so a rank that contains multiple teams is automatically given relatively more weight. For example, if we have eight teams that are ranked in four categories as 1st place, 2nd place, 3rd--4th, and 5th-8th place then the relative weights would be 1, 1, \(\frac12\), \(\frac14\), respectively, which would translate into actual weights
\(w=(\frac{6}{5}, \frac{6}{5}, \frac{3}{5})\) since the sum of the weights for the first \(3\) ranks should be \(3\).

To assign a weighting scheme to the prediction intervals, we may borrow the ``doubling'' concept utilized in NCAA basketball (\citet{espn}). Here, the last interval containing the teams with the worst finishing places, which is also the largest interval, receives a weight of 1. From there, each ensuing interval receives a weight which is double that of its predecessor. Therefore, in the context of a 32-team tournament such as the World Cup, the \emph{relative} weights for the intervals will be: 16, 8, 4, 2, and 1. Again, these relative weights would translate in actual weights \(w=(\frac{32}{15}, \frac{16}{15}, \frac{8}{15}, \frac{4}{15})\), which places emphasis on prediction accuracy in the final match, and a relatively low importance on predictions for the worst-performing teams in the tournament.

In 2026, the FIFA World Cup will be introducing a new format with an expanded pool of 48 teams. The teams will be split up into 16 groups consisting of 3 teams each. The last-placed team in each group will be eliminated, and the rest will move on to the ``Round of 32'', where teams will play a knockout format until a single winner is found. The aforementioned weighted scoring system can be implemented all the same, where the lowest interval, composed of teams who fail to advance out of the group stage, receives a relative weight of 1, and each subsequent interval successively doubles the weight. Thus, the complete scoring system for the 2026 World Cup tournament consisting of 48 teams could use relative weights of 32, 16, 8, 4, 2, and 1.

\hypertarget{simulations}{%
\section{Simulations}\label{simulations}}

We used simulations to investigate the TRPS as a performance measure of tournament predictions. For the simulations we assumed a simple Bradley-Terry model \citep{bradley1952}, which sets the chance for team A to win against team B to
\[\frac{\beta_A}{\beta_A+\beta_B}.\]
where \(\beta_A\) and \(\beta_B\) reflect the strengths of teams A and B, respectively. In the simulations we compare three different tournament types that are often used in various sports tournaments: a knockout tournament, a single round robin tournament, and a double round robin tournament.

\begin{itemize}
  \item For the knockout tournaments, we assume that we have $N$ rounds, $2^N$ teams and $N+1$ different rank categories. A team proceeds to round $n+1$ if it wins against its opponent in round $n$. For example, we can have $N=4$ such that there are $16$ teams and $5$ different ranks (eight finals, quarter finals, semi-finals, final, winner). This setup reflects many tournaments such as Grand Slams in tennis and also the second part of the FIFA World Cup.
  \item In the single round robin tournament, each team plays each other team once. When a team wins it gets one point and otherwise it gets none. We disregard the possibility of a draw and when two teams have equal points, the rank is determined randomly. This type of tournament resembles the group stages of the FIFA World Cup.
  \item The double round robin tournament is set up similarly to the single round robin tournament, but now each team plays twice against each of the other teams. This is the type of tournament seen for example in the English Premier League or the German Bundesliga.
\end{itemize}

In our simulations we sampled the strengths \(\beta_1, \ldots,\beta_T\) of the \(T\) teams in the competition from a log-normal distribution with shape parameter \(\sigma\). We vary \(\sigma\), which represents the spread of the team strengths, and the number of teams in each tournament (considering sizes of 8, 16, or 32).

We compare the TRPS for the following three generic tournament predictions:

\begin{description}
  \item[True strength prediction:] We assume that the true strength of each team is known and use  these latent strengths to estimate the probabilities that a team will attain a specific rank. In practice, these true probabilities are derived from the Bradley-Terry model by taking the mean of $S=10000$ simulations using the true team strengths in order to count how often a team reaches a given rank.
  
   The true strength prediction reflects the optimal prediction we could make prior to the start of the tournament. 
  \item[Flat prediction:] Each team has equal chance to reach any of the ranks.
  \item[Confident prediction:] For the confident prediction we know the ordering of the true team strengths but we do not know their relative difference. The team with the highest strength will be given a 100\% chance of obtaining rank 1, the team with the second highest strength will be a certain prediction of rank 2 etc. Consequently, we assume that the order of the teams will provide a perfect prediction of the final results. 

\end{description}

\begin{table}[ht]
\centering
\begin{tabular}{lrrlrlrr}
  \hline
Tour-   & & & \multicolumn{3}{c}{TRPS} & \\ 
nament   & & & \multicolumn{3}{c}{(average $\pm$ standard deviation)} & \\ \cline{4-6}
type & Teams & $\sigma$ & True & Flat & Conf & P(TSP $<$ FP) & P(TSP $<$ CP) \\ 
  \hline
\multirow{9}{*}{Knockout} &   8 &   1 & 0.129$ \pm $0.048 & 0.18 & 0.255$ \pm $0.065 & 0.85 & 1.00 \\ 
   &  16 &   1 & 0.111$ \pm $0.026 & 0.15 & 0.188$ \pm $0.059 & 0.92 & 0.98 \\ 
   &  32 &   1 & 0.089$ \pm $0.019 & 0.13 & 0.165$ \pm $0.03 & 0.96 & 1.00 \\ 
   &   8 &   2 & 0.083$ \pm $0.066 & 0.18 & 0.204$ \pm $0.062 & 0.91 & 1.00 \\ 
   &  16 &   2 & 0.083$ \pm $0.027 & 0.15 & 0.141$ \pm $0.046 & 0.98 & 1.00 \\ 
   &  32 &   2 & 0.062$ \pm $0.019 & 0.13 & 0.123$ \pm $0.019 & 1.00 & 1.00 \\ 
   &   8 &   3 & 0.04$ \pm $0.04 & 0.18 & 0.132$ \pm $0.056 & 0.99 & 0.97 \\ 
   &  16 &   3 & 0.055$ \pm $0.027 & 0.15 & 0.115$ \pm $0.039 & 0.99 & 1.00 \\ 
   &  32 &   3 & 0.038$ \pm $0.015 & 0.13 & 0.098$ \pm $0.016 & 1.00 & 1.00 \\ 
   \\
  \multirow{9}{*}{\shortstack{Single\\round\\robin}} &   8 &   1 & 0.098$ \pm $0.035 & 0.19 & 0.138$ \pm $0.06 & 0.99 & 0.90 \\ 
   &  16 &   1 & 0.075$ \pm $0.016 & 0.18 & 0.11$ \pm $0.027 & 1.00 & 0.99 \\ 
   &  32 &   1 & 0.059$ \pm $0.009 & 0.17 & 0.087$ \pm $0.015 & 1.00 & 1.00 \\ 
   &   8 &   2 & 0.082$ \pm $0.028 & 0.19 & 0.12$ \pm $0.052 & 1.00 & 0.89 \\ 
   &  16 &   2 & 0.058$ \pm $0.011 & 0.18 & 0.089$ \pm $0.022 & 1.00 & 0.99 \\ 
   &  32 &   2 & 0.044$ \pm $0.006 & 0.17 & 0.064$ \pm $0.01 & 1.00 & 1.00 \\ 
   &   8 &   3 & 0.049$ \pm $0.024 & 0.19 & 0.066$ \pm $0.039 & 1.00 & 0.77 \\ 
   &  16 &   3 & 0.047$ \pm $0.011 & 0.18 & 0.069$ \pm $0.02 & 1.00 & 0.97 \\ 
   &  32 &   3 & 0.036$ \pm $0.006 & 0.17 & 0.053$ \pm $0.009 & 1.00 & 1.00 \\ 
  \\
  \multirow{9}{*}{\shortstack{Double\\round\\robin}} &   8 &   1 & 0.083$ \pm $0.031 & 0.19 & 0.119$ \pm $0.054 & 0.99 & 0.87 \\ 
   &  16 &   1 & 0.064$ \pm $0.015 & 0.18 & 0.094$ \pm $0.025 & 1.00 & 0.99 \\ 
   &  32 &   1 & 0.052$ \pm $0.007 & 0.17 & 0.077$ \pm $0.013 & 1.00 & 1.00 \\ 
   &   8 &   2 & 0.056$ \pm $0.024 & 0.19 & 0.076$ \pm $0.042 & 1.00 & 0.83 \\ 
   &  16 &   2 & 0.041$ \pm $0.009 & 0.18 & 0.061$ \pm $0.018 & 1.00 & 0.97 \\ 
   &  32 &   2 & 0.034$ \pm $0.005 & 0.17 & 0.051$ \pm $0.009 & 1.00 & 1.00 \\ 
   &   8 &   3 & 0.038$ \pm $0.02 & 0.19 & 0.052$ \pm $0.034 & 1.00 & 0.77 \\ 
   &  16 &   3 & 0.03$ \pm $0.008 & 0.18 & 0.045$ \pm $0.016 & 1.00 & 0.93 \\ 
   &  32 &   3 & 0.025$ \pm $0.004 & 0.17 & 0.037$ \pm $0.007 & 1.00 & 1.00 \\ 
   \hline
\end{tabular}
\caption{Average TRPS (and SD) for various combinations of tournament type, number of teams, team strength variance, and prediction model based on $10000$ simulations. The final two columns show the probabilities that the optimal prediction (TSP = true strength prediction) yields a better (smaller) TRPS than both a flat (FP = flat prediction) and a confident prior-rank-based prediction (CP = confident prediction).  }
\end{table}

The summary of the behaviour of the TRPS for \(10000\) simulated predictions for each combination of tournament type, number of teams, team strength variances and prediction method are shown in Table 1.
The table clearly shows some trends in the behaviour of the TRPS:

\begin{itemize}
  \item The larger the number of rounds (and consequently also ranks and matches) the lower the average TRPS for all three generic predictions. This is similar to what we saw in Figure 2.
  \item The mean TRPS depends on the tournament structure. Tournament structures with more matches (the single and double round robin tournament types) will lead to a lower TRPS for both the true strength predictions and the confident predictions. This is as expected as larger number of matches will allow the team rankings to converge to a more stable result of their relative strengths.   
  \item The larger the spread of the team strengths (higher $\sigma$), the lower the TRPS for the true strength prediction and the confident predictions. This is also as expected, since larger variation in team strengths will make it easier to discriminate between the teams.
    \item The larger the number of rounds/ranks/teams, the better the TRPS is able to distinguish good predictions (close to the true probabilities) from flat predictions.
\end{itemize}

It is also worth noting that the confident predictions based on ranks fare remarkably bad in knockout tournaments when the variance of the team strengths is low. This is perhaps not surprising since the teams are more or less equal in strength and it is harder to determine an obvious winner. However, the confident prediction on average performs \emph{worse} than a simple flat prediction. Making confident (but wrong) predictions has a price!

Recall that the primary purpose of the TRPS is to compare different predictions for one tournament rather than compare a single prediction across different tournaments. While the TRPS can be used do discriminate between predictions we can also use the TRPS to combine the predictions into an ensemble predictor as indicated in the next section.

\hypertarget{improving-predictions-using-ensemble-predictions}{%
\section{Improving predictions using ensemble predictions}\label{improving-predictions-using-ensemble-predictions}}

Typically, there is not a single ``best'' model to predict the output of a tournament, and different statistical models may capture different aspects of the prediction. Combining information
from several models can improve the overall prediction by pooling results from
multiple prediction models into a single ensemble prediction. Bayesian
model averaging is a standard technique to generate an ensemble
prediction model by combining different statistical models and taking their uncertainty into account. It works by creating a weighted average of the predictions from each individual
model where the weight assigned to each model is given by how well it performed prior to the prediction \citep{bma1, bma2}.

Let us assume that we have \(K\) prediction models, \(M_1, \ldots, M_K\), and for each model \(M_k\) we have a corresponding prediction \(X^k\). Then the ensemble Bayesian model average prediction, \(X^\star\), is given as the weighted average

\[X^\star = \sum_{k=1}^K X^k P(M_k|y),
\]
where \(P(M_k|y), k=1, \ldots, K\), are the weights of the \(K\) models based on prior training data \(y\). The weights \(P(M_k|y)\) depend on the prior distribution of the models and may be difficult to compute. However, the weights of the individual prediction models can be estimated directly if predictions (and final results after the tournaments have ended) based on the \emph{same} models are available from prior tournaments. Here we will sketch how the TRPS can be used in combination with ensemble prediction.

Let \(\tilde{X}^k\) represent the prediction from model \(M_k\) at a previous tournament and let \(\widetilde{O}\) be the actual outcome of the previous tournament. Then

\[\hat\omega = \text{arg min}_{\omega_1, \ldots, \omega_K; \sum_{k=1}^K \omega_k=1} \; \text{TRPS}(\widetilde{O}, \sum_{k=1}^K  \omega_k \tilde{X}^k)\]

\noindent will be an optimal combination of weights based on the previous predictions. An ensemble prediction for a new tournament would then be
\[
X^\star = \sum_{k=1}^K \hat\omega_k X^k
\]
and this ensemble prediction will reduce both bias and variance of the final prediction for the prior tournament and may also provide a better prediction for the current \citep{Jin2006}. However, since the weights are clearly optimized to match the quality of the predictions in the previous tournament, they may be overfitted to \(\widetilde{X}^k\) and \(\widetilde{O}\) and not optimal for the current prediction.

If we have information on predictions and outcomes from the same set of models from several prior tournaments, then we can use the average TRPS based on the prior tournaments to optimize the weights we use for the current prediction. By employing several prior tournaments we will reduce the risk of overfitting since the number of weights will remain constant. Specifically, if we have predictions from the same set of \(K\) prediction models from \(J\) prior tournaments then the optimal weights can be computed as

\[\hat\omega = \text{arg min}_{\omega_1, \ldots, \omega_K; \sum_{k=1}^K \omega_k=1} \; \frac{1}{J} \sum_{j=1}^J \text{TRPS}(\widetilde{O_j}, \sum_{k=1}^K  \omega_k \tilde{X}_j^k),\]

\noindent  where \(\widetilde{O_j}\) and \(\tilde{X}_j^k\) respectively stand for the actual outcome and prediction from prediction model \(M_k\) in tournament \(j=1,\ldots,J\).

\hypertarget{worldcup}{%
\section{Application: Evaluating predictions for the 2018 FIFA World Cup}\label{worldcup}}

For the 2018 FIFA World Cup competition there were 32 teams, initially split into 8 groups. An online prediction tournament was made public and analysts were encouraged to contribute predictions before the World Cup started on June 14th, 2018 (\citet{fifa20182}, \citet{fifa20183}, \citet{fifa20181}). Each prediction competition contestant had to submit a 32x32 matrix as in Section \ref{trps}, and the rows of each prediction were subsequently collapsed into the 7 actual possible rank categories that we are able to observe from the tournament: 1st place, 2nd place, 3rd place, 4th place, 5th-8th place, 9th-16th place, and 17th-32nd place. These seven categories comprised 1, 1, 1, 1, 4, 8, and 16 teams, respectively.

As part of the competition it was announced that a weighted log loss penalty would be used for scoring the individual predictions (\citet{logloss}). The log loss uses the average of the logarithm of the prediction probabilities and is given by

\[-\frac{1}{T}\sum_{t=1}^T\sum_{r=1}^R w_r \log(X_{rt}) \mathbbm{1}(O_{rt} = 1),
\]

where \(w_r\) was a pre-specified vector of weights given to each team within each of the seven possible collapsed rank categories, \(w_r = (1, 1, \frac12, \frac12, \frac14, \frac18, \frac1{16})\).
\(\mathbbm{1}(O_{rt}=1)\) is the indicator function of a correctly predicted rank for each team.
The log loss has the nice property that it places large penalties on confident predictions that are wrong. However, it is not obvious how to handle the log loss when a prediction of 0\% has been made (and it occurs) and it also uses the same penalty to predictions that were almost the correct rank as it does to predictions that were far apart in ranks.

Here we will focus on the individual predictions, how they compare to each other and to the final tournament result when we use the TRPS and the wTRPS. For completion we will also include the scores for the original log loss.

Five predictions entered the contest:

\begin{enumerate}
\def\labelenumi{\arabic{enumi}.}
\tightlist
\item
  A completely flat and trivial prediction where every team had the same probability \(\frac{1}{32}\) of obtaining each of the original 32 ranks. Since these 32 ranks - one for each team - are collapsed to the seven possible ranks we can observe, the flat probability prediction probabilities become
  \(\frac{1}{32}, \frac{1}{32}, \frac{1}{32}, \frac{1}{32}, \frac{4}{32}, \frac{8}{32}, \frac{16}{32}\) for ranks 1--7, respectively.
\item
  A prediction by Ekstrøm using a Skellam distribution to model the difference in goals scored by each team in a match depending on the teams' initial skill level derived from a betting company \citep[\citet{skellam2}, \citet{fifa20181}, \citet{fifa20182}, \citet{fifa20183}]{skellam}.
\item
  Another prediction by Ekstrøm using the Bradley-Terry model using each team's ELO ranking as the team strength \citep{bradley1952}. The ELO rankings were downloaded from \texttt{https://www.eloratings.net/} immediately prior to the start of the World Cup.
\item
  A prediction by \citet{groll2018} where a random forest model was used to estimate team strengths based on several team-specific covariates (e.g., mean age, number of players from strong clubs, economic factors, coach factors), including as most important covariate individual team ability parameters that reflect a team's current strength based on historic matches.
\item
  An updated version of the prediction by \citet{groll2018} has been provided after the World Cup and is actually the content of the journal paper \citet{groll2019}. This prediction was submitted after the competition ended but is included here for the sake of completion. Besides small algorithmic improvements, it added the ELO ranking to the predictors in the random forest.
\end{enumerate}

Table 2 below shows the tournament rank probability scores for the five predictions. For the wTRPS we have used weights similar to the weights for the log loss described above except that the vector was scaled to sum to 6 (one less than the number of rank categories).

\begin{table}

\caption{\label{tab:unnamed-chunk-4}TRPS, wTRPS and log loss scores for 5 predictions from the recent 2018 FIFA World Cup. Smaller numbers indicate better predictions. The scale of the numbers differs between the different methods, and cannot be directly compared.}
\centering
\begin{tabular}[t]{lrrr}
\toprule
  & TRPS & wTRPS & Log loss\\
\midrule
\rowcolor{gray!6}  Flat & 0.120 & 0.214 & 0.455\\
Ekstrøm (Skellam) & 0.086 & 0.153 & 0.367\\
\rowcolor{gray!6}  Ekstrøm (ELO) & 0.101 & 0.179 & 0.421\\
Groll et al. (2018) & 0.089 & 0.157 & 0.365\\
\rowcolor{gray!6}  Groll et al. (2019) & 0.090 & 0.159 & 0.371\\
\bottomrule
\end{tabular}
\end{table}

Not surprisingly, the flat prediction fares worst and obtains a TRPS of 0.120. The best of the five predictions is Ekstrøm's Skellam-based model which uses odds from a Danish bookmaker as input and it obtains a TRPS of 0.086. The model by \citet{groll2018} has a TRPS almost similar to the Skellam model. Interestingly, the updated model that was provided after the tournament performs slightly worse than the original model by \citet{groll2018} although the differences are negligible. The weighted tournament rank probability score, wTRPS, provides essentially the same ranking of the five predictions, whereas the log loss swaps the order of the two best predictions. This swap is due to the fact that the Ekstrøm (Skellam) prediction placed markedly high probabilities on expected favorites such as Germany, Brazil and Spain, all of which did not fare so well at the World Cup.

The most surprising result is the difference obtained in scores between the ELO-based model by Ekstrøm and the corresponding Skellam-based model. While the two models are slightly different, their main differences are due to the input information that is given to the model. For the Skellam model the input are bookmaker odds while the ELO-based model uses the official ELO ranking of the teams. Perhaps the ELO rating is not updating fast enough to reflect the current quality of the teams entering the World Cup --- which is something the bookmakers more readily can adjust for.

The average of the two best predictions yields a TRPS score of 0.085 which is a slight improvement in TRPS. This suggests that it might be beneficial to create an ensemble predictor for future predictions.

\hypertarget{discussion}{%
\section{Discussion}\label{discussion}}

In this paper we have extended the rank probability score and introduced the tournament rank probability score to create a measure to evaluate and compare the quality of pre-tournament predictions. The TRPS is a proper scoring rule and it retains the desirable properties of the rank probability score --- being non-local and sensitive to distance --- such that predictions that are almost right are more desirable than predictions that were clearly wrong.

Our simulations show that the TRPS is very flexible, handles partial rankings, works well with different tournament types and is able to capture, evaluate and rank predictions that are more than random guesses. As such it sets the basis for future discussions and evaluations of predictions that appear in research and media, especially in situations where prior information may be severely limited, in the sense that two teams may have never met faced each other in a match before.

The flexibility of the TRPS comes in two guises. First, it is possible to set specific weights to use the TRPS to focus on predicting specific results such as the winner. This ensures that the same framework can be used for general tournament predictions as well as specialized forecasts.
Secondly, the TRPS is directly applicable to partial rankings which is common in many tournaments.
In fact, instead of adding weights, one could choose to take another way of weighing by collapsing the relevant categories into one. For example, in the Premier League we could disregard the normal ranking of the teams from 1 to 20, and instead only consider the ranks 1 (champion), 2-4 (qualified for Champions league), 5 (qualified for Europa League), 6-16 and 17-20 (degradation). In Olympic games, the most important ranks are of course the first three places. Next to this, the first 8 competitors receive an ``Olympic Diploma''. So the ranks here could be Gold (1), Silver (2), Bronze (3), Olympic Diploma (4-8), and Others (9+).

The TRPS leads naturally to creating improved model-averaged ensemble predictions that combine a set of prediction models into an optimal prediction based on results from earlier tournaments. We have outlined how the TRPS can be used for this purpose and we will pursue this in a future publication.

In conclusion, the TRPS provides a general measure to evaluate the quality and precision of tournament predictions within many fields and it can serve as a way to both determine prediction winners and provide the foundation for creating ensemble model-averaged predictions that increase the overall precision of sport tournament predictions within all fields of sport.

\bibliographystyle{tfcad}
\bibliography{bibliography.bib}

\end{document}